\documentclass{iopjournal}
\usepackage{mathrsfs}
\usepackage{amsfonts}
\usepackage{amsmath}
\usepackage{amssymb}
\usepackage{graphicx,subfigure}
\usepackage{bm}
\usepackage{color}
\usepackage[normalem]{ulem}
\usepackage{dcolumn} 
\usepackage{float}
\usepackage{tabularx}
\usepackage{hyperref}
\newcommand{\ket}[1]{|#1\rangle}
\newcommand{\bra}[1]{\langle #1|}

\begin{document}

\articletype{Paper}

\title{{\it Quasi}-holonomy in non-adiabatic quantum evolution} 

\author{Erik Sj\"oqvist$^{1,*}$ and Adam Fredriksson$^2$}

\affil{$^1$Department of Physics and Astronomy, Uppsala University, 
Box 524, SE-751 20 Uppsala, Sweden}

\affil{$^2$Naturwissenschaftlich-Technische Fakult\"at, Universit\"at Siegen, Walter-Flex-Stra{\ss}e 3, D-57068 Siegen, Germany}

\affil{$^*$Author to whom any correspondence should be addressed.}

\email{erik.sjoqvist@physics.uu.se}


\date{\today}

\begin{abstract}
We develop a framework for {\it quasi}-holonomy in non-adiabatic quantum time evolution of subspaces along loops in a complex Grassmannian. By factoring the Schr\"odinger evolution into dynamical and connection-induced contributions in a moving basis, we obtain an effective geometric generator that depends explicitly on the dynamical propagator. This {\it quasi}-connection does not define a genuine connection on the original Grassmann bundle, since its gauge transformation law acquires a history-dependent, nonlocal term. Other ways of  factoring the Schrödinger evolution are briefly discussed. All these approaches suffer from the same type of history-dependence,  thereby defining transport of subspaces in which geometric and dynamical effects are generally intertwined, just as in the case of the {\it quasi}-holonomy. Our work sheds light on the issue of separating quantum evolution of subspaces into holonomic and dynamical parts from an essentially gauge-theoretic perspective. 
\end{abstract}


\section{Introduction}
The geometry of quantum time evolution of subspaces of state space is the geometry of paths in Grassmannian fibre bundles. Such evolution involves two conceptually distinct structures: a holonomic contribution associated with the Grassmannian connection and a dynamical contribution generated by the Hamiltonian. The dynamical part accounts for the accumulated action of the Hamiltonian along the path, 
whereas the holonomic part reflects the twisting of the evolving subspace induced by the curved geometry of the underlying Grassmannian bundle. This geometric twisting is encoded in a matrix-valued geometric phase, or holonomy, arising from the gauge structure of the bundle and generated by its non-Abelian connection \cite{wilczek84,anandan88}.

A central question in the experimental context is whether these two aspects of the evolution can be separated in a meaningful way. While such a 
separation is possible in special situations \cite{fredriksson26b}, it is generally obstructed by the noncommutativity of the holonomic and dynamical generators, causing the corresponding contributions to become intrinsically intertwined during the evolution. The purpose of the present work is to put this inseparability feature on a firm gauge-theoretic footing, by systematically  factoring the Schr\"odinger evolution with respect to either the holonomic or dynamical propagator. This reveals a class of history-dependent effective generators that clarify the extent to which geometric and dynamical effects can be distinguished.

\section{Theory of {\it quasi}-holonomy}
\subsection{Subspace evolution}
Let $C: t \in [0,\tau] \mapsto \mathscr{V}_M (t)$ be a loop\footnote{Although the formalism extends naturally to open paths in the Grassmannian, their treatment introduces notational and conceptual subtleties, including the possibility of partial holonomies \cite{kult06}. Since these issues are not central to the present work, we restrict our analysis to cyclic evolution for clarity.} of $M$-dimensional 
subspaces of $N$-dimensional Hilbert space, i.e., $C$ is a loop in the complex Grassmannian 
$\mathscr{G} (N;M)$. Physically, we assume $C$ is generated by the Schr\"odinger 
equation with Hamiltonian $H (t)$, yielding the matrix differential equation 
\cite{anandan88}
\begin{eqnarray}
\dot{\mathbf{W}} (t) = \mathbf{K}(t)\mathbf{W}(t) + \mathbf{A}(t)\mathbf{W}(t) , 
\label{eq:ae}
\end{eqnarray}
where $\mathbf{A}_{jk} (t) = \langle \dot{\phi}_j (t) \ket{\phi_k (t)}$ and $\mathbf{K}_{jk} (t) = 
-i\bra{\phi_j (t)} H (t) \ket{\phi_k (t)}$ are the holonomy and dynamical matrix generators, 
respectively. Here, $\{ \ket{\phi_j (t)} \}_{j = 1}^{M}$ is a local section spanning the $M$-dimensional subspace 
$\mathscr{V}_M (t)$, such that $\langle \phi_j (0) \ket{\phi_k (\tau)} = \delta_{jk}$. 
The formal solution of Eq.~\eqref{eq:ae} is
\begin{eqnarray}
\mathbf{W} (\tau) = \mathcal{T} e^{\int_0^{\tau} [\mathbf{K}(t)+\mathbf{A}(t)] dt},
\label{eq:aesol}
\end{eqnarray}
where $\mathcal{T}$ denotes time-ordering. Here, we have assumed the initial condition $\mathbf{W} (0) = \mathbf{I}_M$, $\mathbf{I}_M$ being the $M\times M$ identity matrix. The unitary propagator $\mathbf{W} (\tau)$ connects a Schr\"odinger solution $\ket{\psi (t)} \in \mathscr{V}_M (t)$ at $t=\tau$ and the local section, according to 
\begin{eqnarray}
\ket{\psi (\tau)} = \sum_{k=1}^M 
\ket{\phi_k (0)} \mathbf{W}_{kj} (\tau) , 
\end{eqnarray}
assuming $\ket{\psi (0)} = \ket{\phi_j (0)}$ for some $j=1,\ldots,M$. In the language of quantum computation, $\mathbf{W} (\tau)$ more generally defines a unitary operator 
\begin{eqnarray}
U & = & \sum_{j,k=1}^M \mathbf{W}_{jk} (\tau) \ket{\phi_j (0)} \bra{\phi_k (0)} , 
\end{eqnarray}
being a quantum gate acting on superpositions of the computational states $\{ \ket{\phi_j (0)} \}_{j=1}^M$ 
that span the $M$-dimensional computational subspace of the system. 

\subsection{Definition of quasi-holonomy}
Now, define $\mathbf{D} (t)$ to be the solution of the dynamical part of 
Eq.~\eqref{eq:ae}, i.e., 
\begin{eqnarray}
\dot{\mathbf{D}} (t) = \mathbf{K} (t) \mathbf{D} (t) ,
\label{eq:dyn}
\end{eqnarray}
with $\mathbf{D} (0) = \mathbf{I}_M$. Formally, 
\begin{eqnarray}
\mathbf{D} (\tau) = \mathcal{T}
e^{\int_0^{\tau} \mathbf{K} (t) dt}.
\end{eqnarray}
By taking the derivative of the ansatz 
\begin{eqnarray}
\mathbf{W}(t) = \mathbf{D} (t) \mathbf{V} (t)
\label{eq:ansatz}
\end{eqnarray}
and using Eqs.~~\eqref{eq:dyn} and \eqref{eq:ae}, we find 
\begin{eqnarray}
\dot{\mathbf{W}} (t) = 
\dot{\mathbf{D}} (t) \mathbf{V} (t) + \mathbf{D} (t) \dot{\mathbf{V}} (t) 
& = &  \mathbf{K} (t) \mathbf{W} (t) + \mathbf{D} (t) \dot{\mathbf{V}} (t) ,
\nonumber \\ 
 & = & 
\mathbf{K} (t) \mathbf{W} (t) + \mathbf{A} (t) \mathbf{D} (t) \mathbf{V} (t),  
\end{eqnarray}
which implies 
\begin{eqnarray}
\dot{\mathbf{V}} (t) = 
\mathbf{D}^{\dagger} (t) \mathbf{A} (t) \mathbf{D} (t) \mathbf{V} (t) . 
\label{eq:ip}
\end{eqnarray}
Upon integration, we obtain the {\it quasi}-holonomy: 
\begin{eqnarray}
\mathbf{V} (\tau) = \mathcal{T} e^{\int_0^{\tau}
\mathbf{D}^{\dagger} (t) \mathbf{A} (t) \mathbf{D} (t) dt},
\end{eqnarray}
by using the initial condition $\mathbf{V}(0) = \mathbf{D}^{\dagger} (0) \mathbf{W} (0) = \mathbf{I}_{M}$.
The full solution of Eq.~\eqref{eq:ae} can thus be written as\footnote{The ordering of the factors on the right-hand side of Eqs.~\eqref{eq:dyn} and \eqref{eq:ansatz} is crucial. As an illustration of this point, consider the reversed ordering $\dot{\tilde{\mathbf{D}}} (t) = \tilde{\mathbf{D}}(t) \mathbf{K} (t)$ and $\mathbf{W} (t) = \tilde{\mathbf{V}}(t) \tilde{\mathbf{D}} (t)$ with $\tilde{\mathbf{D}} (0) = \tilde{\mathbf{V}} (0) = \mathbf{I}_M$. By taking the derivative of the latter and using Eq.~\eqref{eq:ae}, we find 
$\dot{\tilde{\mathbf{V}}} (t) \tilde{\mathbf{D}} (t) + \tilde{\mathbf{V}} (t) \tilde{\mathbf{D}} (t) \mathbf{K} (t) = [\mathbf{A} (t) + \mathbf{K} (t)] \tilde{\mathbf{V}} (t) \tilde{\mathbf{D}} (t)$. Rearranging and multiplying with $\tilde{\mathbf{D}}^{\dagger} (t)$ from the right,  yields
\begin{eqnarray*}
\dot{\tilde{\mathbf{V}}} (t) = [\mathbf{A} (t) + \mathbf{K} (t)] \tilde{\mathbf{V}} (t) - \tilde{\mathbf{V}} (t) \tilde{\mathbf{D}} (t) \mathbf{K} (t)\tilde{\mathbf{D}}^{\dagger} (t) . 
\end{eqnarray*}
To proceed from here, note that given operators $\mathbf{L}(t)$, $\mathbf{R}(t)$ satisfying the differential equations $\dot{\mathbf{L}}(t) = \mathbf{B}(t)\mathbf{L}(t)$ and $\dot{\mathbf{R}}(t) = \mathbf{R}(t)\mathbf{C}(t)$, the solutions of which being a time-ordered and reverse time-ordered exponential, respectively, the operator $\mathbf{U}(t) = \mathbf{L}(t)\mathbf{R}(t)$ obeys the differential equation $\dot{\mathbf{U}}(t) = \mathbf{B}(t)\mathbf{U}(t) + \mathbf{U}(t)\mathbf{C}(t)$. Thus, 
\begin{eqnarray*}
\mathbf{W} (\tau) = \mathbf{W} (\tau) \bar{\mathcal{T}} e^{-\int_0^{\tau} \tilde{\mathbf{D}} (t)\mathbf{K} (t)\tilde{\mathbf{D}}^{\dagger} (t) dt} \tilde{\mathbf{D}} (\tau) 
\end{eqnarray*}
with $\bar{\mathcal{T}}$ reverse time-ordering and we have used Eq.~\eqref{eq:aesol}. This is a trivial result that only confirms the solution of the reversed ordered equation $\dot{\tilde{\mathbf{D}}} (t) = \tilde{\mathbf{D}}(t) \mathbf{K} (t) = [\tilde{\mathbf{D}}(t) \mathbf{K} (t) \tilde{\mathbf{D}}^{\dagger} (t)] \tilde{\mathbf{D}}(t)$ but give no information about the full solution $\mathbf{W} (\tau)$. 
\label{footnote2} 
}
\begin{eqnarray}
\mathbf{W}(\tau) = \mathcal{T} e^{\int_0^{\tau} \mathbf{K} (t) dt} 
\mathcal{T} e^{\int_0^{\tau} \mathbf{A}_D  (t) dt} ,
\label{eq:factor1}
\end{eqnarray}
where we have defined the non-Abelian (matrix-valued) {\it quasi}-connection: 
\begin{eqnarray}
\mathbf{A}_D  (t) \equiv \mathbf{D}^{\dagger} (t) \mathbf{A} (t) \mathbf{D} (t) . 
\label{eq:qc}
\end{eqnarray}
Note that both factors in Eq.~\eqref{eq:factor1} are time-ordered in forward direction. 

\subsection{Properties of {\it quasi}-holonomy}
It is essential to realize that Eq.~\eqref{eq:factor1} does not represent 
a separation into a holonomic and dynamical part, as 
the second factor depends explicitly on the first factor $\mathbf{D} (t)$. This has some 
serious consequences for the `holonomic nature' of the {\it quasi}-holonomy. To see this, let us 
consider how the {\it quasi}-connection transforms under a time-dependent change of local section 
(gauge transformation): 
\begin{eqnarray}
\ket{\phi_j(t)} \mapsto \ket{\phi'_j(t)} = \sum_{k = 1}^{M} \ket{\phi_k(t)} \mathbf{\Omega}_{kj}(t),
\label{eq:gt}
\end{eqnarray}
where $\mathbf{\Omega}(t)\in U(M)$ is a smooth unitary matrix such that $\mathbf{\Omega} (\tau) = \mathbf{\Omega} (0) = \mathbf{I}_M$.

While the connection $\mathbf{A}(t)$ transforms as
\begin{eqnarray}
\mathbf{A} (t) \mapsto \mathbf{A}'(t) = \mathbf{\Omega}^{\dagger}(t)\,\mathbf{A}(t)\,\mathbf{\Omega}(t) 
+  \dot{\mathbf{\Omega}}^{\dagger}(t) \mathbf{\Omega} (t) ,
\label{eq:true_gt}
\end{eqnarray}
the dynamical generator $\mathbf{K}(t)$ transforms by conjugation:
\begin{eqnarray}
\mathbf{K} (t) \mapsto \mathbf{K}'(t) = \mathbf{\Omega}^{\dagger}(t)\,\mathbf{K}(t)\,\mathbf{\Omega}(t).
\end{eqnarray}
The dynamical propagator $\mathbf{D}(t)$ transforms as
\begin{eqnarray}
\mathbf{D} (t) \mapsto \mathbf{D}'(t) = \mathbf{\Omega}^{\dagger}(t)\,\mathbf{D}(t) , 
\label{eq:dyn_gt}
\end{eqnarray}
since we have, as stated above, fixed the gauge so that $\mathbf{\Omega}(0) = \mathbf{I}_M$.\footnote{More generally, we have $\mathbf{D} (t) \mapsto \mathbf{D}' (t) = \mathbf{\Omega}^{\dagger} (t) \mathbf{D} (t) \mathbf{\Omega} (0)$, which reduces to Eq.~\eqref{eq:dyn_gt} by assuming $\mathbf{\Omega} (0) = \mathbf{I}_M$.}
Thus, we find that 
\begin{eqnarray}
\mathbf{A}_D  (t) \mapsto \mathbf{A}_D ' (t) & = & \mathbf{D}'^{\dagger}(t) \mathbf{A}'(t) \mathbf{D}'(t)
\nonumber \\
 & = & \left(\mathbf{\Omega}^{\dagger} (t) \mathbf{D} (t) \right)^{\dagger}
\left(\mathbf{\Omega}^{\dagger} (t) \mathbf{A} (t) \mathbf{\Omega}(t)
+  \dot{\mathbf{\Omega}}^{\dagger} (t) \mathbf{\Omega} (t) \right)
\left(\mathbf{\Omega}^{\dagger} (t) \mathbf{D} (t) \right)
\nonumber \\
 & = & \mathbf{A}_D (t) + \mathbf{D}^{\dagger}(t) 
\mathbf{\Omega} (t)  \dot{\mathbf{\Omega}}^{\dagger}(t)\,\mathbf{D} (t),
\label{eq:quasi_gt}
\end{eqnarray}
where we have used Eqs.~\eqref{eq:qc}, \eqref{eq:true_gt}, and \eqref{eq:dyn_gt}. 

The question is whether the {\it quasi}-connection in Eq.~\eqref{eq:qc} should be regarded as a true connection on the Grassmann bundle. As is well-known \cite{chruscinski04,leone19}, a true connection on a principal $U(M)$-bundle must transform under a gauge transformation $\mathbf{\Omega}(t)$ as in Eq.~\eqref{eq:true_gt}, involving a time-local inhomogeneous term $\dot{\mathbf{\Omega}}^{\dagger}(t) \mathbf{\Omega}(t)$ that guarantees covariance of the associated covariant derivative. 

With the connection defined right below Eq.~\eqref{eq:ae}, the associated covariant 
derivative is the time-derivative operator modified so as to act covariantly on the coefficients (probability amplitudes) in the moving basis $\{ \ket{\phi_j(t)} \}_{j=1}^{M}$. Explicitly, for a Schr\"odinger solution 
$\ket{\psi (t)} \in \mathscr{V}_M (t)$, we can write
\begin{eqnarray}
|\psi(t)\rangle = \sum_{j=1}^M \ket{\phi_j(t)} c_j(t) . 
\end{eqnarray}
Under the $U(M)$ gauge transformation given in Eq.~\eqref{eq:gt}, the physical state $\ket{\psi (t)}$ should remain unchanged. Thus, 
\begin{eqnarray}
\ket{\psi (t)} = \sum_{j=1}^M \ket{\phi_j' (t)} c_j' (t) 
= \sum_{j,k=1}^M \ket{\phi_k (t)}
\mathbf{\Omega}_{kj} (t) c_j' (t) ,
\end{eqnarray}
which implies 
\begin{eqnarray}
c_j'(t) = \sum_{l=1}^M \mathbf{\Omega}_{jl}^{\dagger} (t) c_l (t) , 
\label{eq:coeff_gt}
\end{eqnarray}
because $\mathbf{\Omega} (t) \in U(M)$. 

A covariant derivative 
\begin{eqnarray}
\mathcal{D}_t c_j(t) = \frac{d}{dt} c_j(t) - \sum_{k=1}^M \mathbf{A}_{jk} (t) c_k (t)
\end{eqnarray} 
is characterized by the requirement that it transforms in the same way as $c_j (t)$: 
\begin{eqnarray}
\mathcal{D}_t c_j (t) \mapsto 
\mathcal{D}_t' c_j'(t) =  \sum_{k=1}^M \mathbf{\Omega}_{jk}^{\dagger} (t) \mathcal{D}_t c_k (t) . 
\label{eq:cov_gt}
\end{eqnarray}
To verify that the transformation law for $\mathbf{A} (t)$ indeed guarantees this property, we use Eqs.~\eqref{eq:true_gt} and \eqref{eq:coeff_gt} to find 
\begin{eqnarray}
\mathcal{D}_t' c_j'(t ) & = & \frac{d}{dt} c_{j}'(t) - \sum_{k = 1}^{M}\mathbf{A}_{jk}'(t)c_{k}'(t) 
\nonumber \\ 
 & = & 
\sum_{k=1}^M \frac{d}{dt} 
\left[ \mathbf{\Omega}_{jk}^{\dagger} (t) c_k (t) \right] -
\sum_{k,l=1}^M \bigg[ \mathbf{\Omega}^{\dagger} (t) \mathbf{A} (t) \mathbf{\Omega} (t) + 
\dot{\mathbf{\Omega}}^{\dagger} (t) \mathbf{\Omega} (t)
\bigg]_{jk} \mathbf{\Omega}_{kl}^{\dagger} (t) c_l (t) 
\nonumber \\ 
 & = & \sum_{k = 1}^{M}\mathbf{\Omega}_{jk}^{\dagger} (t)\bigg[\frac{d}{dt}c_{k}(t) - \sum_{l = 1}^{M}\mathbf{A}_{kl}(t)c_{l}(t)\bigg] = \sum_{k=1}^M 
\mathbf{\Omega}_{jk}^{\dagger} (t) \mathcal{D}_t c_k (t) , 
\end{eqnarray}
since the inhomogeneous terms cancel. This is precisely why Eq.~\eqref{eq:true_gt} 
is the defining transformation law of a genuine connection: the extra time-local term 
$\dot{\mathbf{\Omega}}^{\dagger} (t) \mathbf{\Omega} (t)$ exactly cancels the derivative acting on the gauge transformation matrix.

By contrast, the {\it quasi}-connection $\mathbf{A}_D  (t)$ does not transform with a time-local inhomogeneous term of the form $\dot{\mathbf{\Omega}}^{\dagger} (t) \mathbf{\Omega} (t)$, but instead acquires the term $\mathbf{D}^{\dagger} (t) \mathbf{\Omega} (t) \dot{\mathbf{\Omega}}^{\dagger}(t) \mathbf{D}(t)$, being local in the sense that it is evaluated at a single instant $t$, but nonlocal in 
the gauge-theoretic sense because it depends on the dynamical propagator
\begin{eqnarray}
\mathbf{D} (t) = \mathcal{T} e^{\int_0^t \mathbf{K} (s) ds},
\label{eq:wm}
\end{eqnarray}
which is a time-ordered functional of the full history of $\mathbf{K} (s)$ on the interval $[0,t]$.

Consequently, the gauge transformation law of the {\it quasi}-connection $\mathbf{A}_D  (t)$ cannot be expressed solely in terms of the instantaneous $\mathbf{A}_D (t)$, $\mathbf{\Omega} (t)$, and 
$\dot{\mathbf{\Omega}} (t)$ at time $t$, but depends explicitly on the past evolution encoded in the dynamical propagator $\mathbf{D}(t)$. In this sense, the transformation of $\mathbf{A}_D  (t)$ fails to satisfy 
the connection transformation law of the Grassmannian principal $U(M)$-bundle. 
In other words, $\mathbf{A}_D  (t)$  does not define a covariant derivative associated 
with a principal $U(M)$-bundle connection on the original Grassmannian bundle. This is the 
precise sense in which $\mathbf{A}_D  (t)$ fails to be a genuine connection on the 
Grassmann bundle.

The {\it quasi}-holonomy is naturally interpreted as an interaction-picture transport law. The propagator $\mathbf{D}(t)$ defines a time-dependent change of frame that removes the explicit dynamical evolution, and $\mathbf{A}_D (t)$ is the corresponding dressed generator governing the remaining transport. 
Geometric and dynamical contributions remain intertwined through the explicit dependence of $\mathbf{A}_D (t)$ on $\mathbf{D}(t)$.

Pure {\it quasi}-holonomy is the special case where the dynamical contribution becomes trivial, 
i.e., $\mathbf{D}(\tau)=\mathbf{I}_M$, and the total evolution reduces to
\begin{eqnarray}
\mathbf{W}(\tau) = \mathcal{T} e^{\int_0^{\tau} \mathbf{A}_D  (t) dt} .
\label{eq:pqh}
\end{eqnarray}
The condition $\mathbf{D} (\tau) = \mathbf{I}_M$ is not equivalent to the {\it quasi}-connection 
reducing to the standard Grassmannian connection. It guarantees only that the final dynamical 
propagator is trivial, whereas the {\it quasi}-connection remains a dynamically dressed object 
depending on the full history of  $\mathbf{D} (t)$. Thus, a purely {\it quasi}-holonomic evolution 
as in Eq.~\eqref{eq:pqh} is logically possible, although, again, it is important to keep in mind 
that the dynamical dressing encoded in $\mathbf{A}_D (t)$ prevents a clean separation 
between geometric and dynamical contributions even in such cases. In other words, the 
transport should be regarded {\it quasi}-holonomic rather than holonomic. In the language of 
quantum computation, it is therefore tempting to define a unitary operator 
\begin{eqnarray}
U_{\rm qh} & = & \sum_{j,k=1}^M \mathbf{V}_{jk} (\tau) \ket{\phi_j (0)} \bra{\phi_k (0)} 
\end{eqnarray}
to be a {\it quasi}-holonomic gate. It is left as an open question whether such gates can be realized under realistic physical conditions while retaining their utility for quantum computation. 

We now address the question under what conditions the {\it quasi}-holonomy connection 
$\mathbf{A}_D  (t)$ can be replaced by $\mathbf{A}(t)$ inside the time-ordered exponential
of the unitary factor $\mathbf{V} (\tau)$. Clearly, this amounts to 
\begin{eqnarray}
[\mathbf{A}(t),\mathbf{D}(t)] = 0, \qquad \forall t \in [0,\tau], 
\label{eq:comm1}
\end{eqnarray}
as it evidently implies 
\begin{eqnarray}
\mathbf{V}(\tau) = \mathcal{T}
e^{\int_0^\tau \mathbf{D}^{\dagger} (t) \mathbf{A}(t) \mathbf{D} (t) dt} = 
\mathcal{T} e^{\int_0^\tau \mathbf{A}(t) dt},
\end{eqnarray}
which is precisely the standard holonomy of the Grassmannian bundle. By using 
Eq.~\eqref{eq:wm}, we may write Eq.~\eqref{eq:comm1} as 
\begin{eqnarray}
\left[ \mathbf{A}(t), \mathcal{T} e^{\int_0^t \mathbf{K}(s)\,ds} \right] = 0 , \qquad \forall t \in [0,\tau] ,
\end{eqnarray}
which essentially means that 
\begin{eqnarray}
[\mathbf{A}(t),\mathbf{K}(s)] = 0, \qquad \forall\, s\leq t, 
\end{eqnarray}
as in this way, $\mathbf{A}(t)$ commutes with every term in the Dyson expansion of
$\mathbf{D}(t)$, implying Eq.~\eqref{eq:comm1}. This is satisfied in the settings that were shown in Ref.~\cite{fredriksson26b} to exhaust all cases where the 
holonomic and dynamical parts separate. These are: (i) $\mathbf{A} (t) = 0$, 
 (ii) $\mathbf{K} (t) = 0$, and (iii) $\mathbf{A} (t) = -i A_j (t) \mathbf{P}_j (0)$ and 
$\mathbf{K} (t) = -i K_j (t) \mathbf{P}_j (0)$, $\mathbf{P}_j (0)$ being mutually orthogonal 
projection matrices ($\mathbf{P}_j (0)\mathbf{P}_k (0) = \delta_{jk} \mathbf{P}_k (0)$). 

Geometrically, the {\it quasi}-holonomy differs from the standard holonomy in the general case, 
because the connection is transported into a dynamically dressed frame according to
$\mathbf{A}(t) \mapsto \mathbf{D}^{\dagger} (t) \mathbf{A}(t) \mathbf{D}(t)$.
The two constructions coincide precisely when $\mathbf{A}(t)$ is invariant under this dynamical transport, or equivalently,
${\rm Ad}_{\mathbf{D}(t)} \bigl( \mathbf{A}(t) \bigr) = \mathbf{A}(t), \forall t \in [0,\tau]$,
meaning that $\mathbf{A}(t)$ lies in the centralizer of $\mathbf{D}(t)$, i.e., it is fixed by the 
adjoint action of $\mathbf{D} (t)$. Under this condition, the dynamical dressing becomes 
invisible and the {\it quasi}-holonomy reduces exactly to the standard Grassmannian holonomy.

\section{Other ways of factoring}
Although the primary purpose of this work is to develop the concept of {\it quasi}-holonomy, it is instructive to compare with other factorizations of the propagator $\mathbf{W}(\tau)$. One such factorization leads to a {\it quasi}-dynamical propagator, obtained by dressing the dynamical generator with the holonomic propagator in direct analogy with the construction of {\it quasi}-holonomy. In addition, there exist two further product representations in which the dressing is performed by the full propagator $\mathbf{W}(t)$ itself. As will be seen below, these latter representations are of a different character: unlike the {\it quasi}-holonomic and {\it quasi}-dynamical constructions, neither of the two constitutes a solution of the original equation of motion, but is rather a rewriting of it.



First, we may define $\mathbf{G} (t)$ to be the solution of the holonomic part of 
Eq.~\eqref{eq:ae}, i.e., 
\begin{eqnarray}
\dot{\mathbf{G}} (t) = \mathbf{A} (t) \mathbf{G} (t) 
\label{eq:hol}
\end{eqnarray}
with $\mathbf{G} (0) = \mathbf{I}_M$. By following the same steps as those leading up to Eq.~\eqref{eq:factor1}, 
one finds that the full solution can be written on the form
\begin{eqnarray}
\mathbf{W}(\tau) = \mathcal{T} e^{\int_0^{\tau} \mathbf{A} (t) dt} 
\mathcal{T} e^{\int_0^{\tau} \mathbf{K}_G (t) dt} , 
\label{eq:factor2}
\end{eqnarray}
where we have defined 
\begin{eqnarray}
\mathbf{K}_G (t) \equiv \mathbf{G}^{\dagger} (t) \mathbf{K} (t) \mathbf{G} (t) , 
\label{eq:dc}
\end{eqnarray}
and, again, both factors in the expression for $\mathbf{W} (\tau)$ being time-ordered in the forward direction. While a true dynamical 
generator $\mathbf{K} (t)$ is time-local, the dressed generator $\mathbf{K}_G (t)$ is highly 
path-nonlocal via its explicit dependence on the holonomy propagator $\mathbf{G} (t)$. Thus, just as in the above case of {\it quasi}-holonomy, the dynamical and geometric contributions are intertwined in $\mathcal{T} e^{\int_0^{\tau} \mathbf{K}_G (t) dt}$, 
which therefore may be called a {\it quasi}-dynamical contribution to the subspace evolution. 

For the other two cases, we may use the unitarity of $\mathbf{W} (t)$ to rewrite 
Eq.~\eqref{eq:ae} as \begin{eqnarray}
\dot{\mathbf{W}} (t) = \mathbf{W} (t) \left[ \mathbf{W}^{\dagger} (t) \mathbf{K}(t) 
\mathbf{W} (t) \right] + \mathbf{A}(t)\mathbf{W}(t)  
\label{eq:ae_mod1}
\end{eqnarray}
or 
\begin{eqnarray}
\dot{\mathbf{W}} (t) = \mathbf{K}(t)\mathbf{W}(t) + 
\mathbf{W} (t) \left[ \mathbf{W}^{\dagger} (t) \mathbf{A}(t) \mathbf{W} (t) \right] . 
\label{eq:ae_mod2}
\end{eqnarray}
The former of these was discussed in Ref.~\cite{yu23}. We may put Eqs.~\eqref{eq:ae_mod1} 
and \eqref{eq:ae_mod2} on the factorized form 
\begin{eqnarray}
\mathbf{W} (\tau) = \mathcal{T} e^{\int_0^{\tau} \mathbf{A} (t) dt} 
\bar{\mathcal{T}} e^{\int_0^{\tau} \mathbf{W}^{\dagger} (t)\mathbf{K} (t)\mathbf{W} (t) dt} 
\label{eq:dynform}
\end{eqnarray}
and 
\begin{eqnarray}
\mathbf{W} (\tau) = \mathcal{T} e^{\int_0^{\tau} \mathbf{K} (t) dt} 
\bar{\mathcal{T}} e^{\int_0^{\tau} \mathbf{W}^{\dagger} (t)\mathbf{A} (t)\mathbf{W} (t) dt} , 
\label{eq:holform}
\end{eqnarray}
respectively, with $\bar{\mathcal{T}}$ being reverse time-ordering. Notice that the time-ordering in different directions in Eqs.~\eqref{eq:dynform} and \eqref{eq:holform} is due to the different ordering of factors in the two terms on the right-hand side of Eqs.~\eqref{eq:ae_mod1} and \eqref{eq:ae_mod2}, see footnote \ref{footnote2}. The second factor in both these expressions involve the dynamical and holonomic generators $\mathbf{K} (t)$ and  $\mathbf{A} (t)$, respectively, but now being dressed by the full solution $\mathbf{W} (t)$. 
Crucially, this dressing by $\mathbf{W} (t)$ means that neither the expression in Eq.~\eqref{eq:dynform} nor Eq.~\eqref{eq:holform} constitutes a way to express the solution of Eq.~\eqref{eq:ae}, but is rather a rewriting of the equation. Moreover, since it is evident from Eq.~\eqref{eq:aesol} that $\mathbf{W} (t)$ is as time-nonlocal as $\mathbf{D} (t)$ and $\mathbf{G} (t)$ above, we conclude that neither Eq.~\eqref{eq:dynform} nor Eq.~\eqref{eq:holform} constitute a separation of the quantum time evolution into holonomic and dynamical parts \cite{fredriksson26a}. In other words, just as in the case of the {\it quasi}-holonomic
and {\it quasi}-dynamical propagators, the geometric and dynamical contributions remain intertwined in these two expressions, in the sense of the original Grassmannian bundle. 


The four factorizations in Eqs.~\eqref{eq:factor1}, \eqref{eq:factor2}, \eqref{eq:dynform}, and \eqref{eq:holform} illustrate different ways of rewriting the subspace evolution, but they are not all of the same character. Equations~\eqref{eq:factor1} and \eqref{eq:factor2} constitute genuine solution formulas: once the dynamical propagator $\mathbf{D}(t)$ or holonomic propagator $\mathbf{G}(t)$ has been determined, the remaining factor is obtained from a closed evolution equation generated by the dressed operators $\mathbf{A}_D(t)$ or $\mathbf{K}_G(t)$. The resulting {\it quasi}-holonomic and {\it quasi}-dynamical propagators therefore provide nontrivial alternative descriptions of the evolution, although geometric and dynamical effects remain intertwined through the dressing. By contrast, Eqs.~\eqref{eq:dynform} and \eqref{eq:holform} do not constitute solutions of the original equation of motion.
Their second factors are dressed by the full solution $\mathbf{W}(t)$ itself, making them implicit identities obtained by rewriting the original equation of motion rather than new solution representations. Thus, only Eqs.~\eqref{eq:factor1} and \eqref{eq:factor2} yield distinct propagators associated with dynamical or holonomic dressing, whereas Eqs.~\eqref{eq:dynform} and \eqref{eq:holform} merely restate Eq.~\eqref{eq:ae} in a different form. All four expressions collapse to the same factorized structure only in the exceptional case where the holonomic and dynamical contributions separate on the underlying Grassmannian bundle, i.e., when the relevant generators commute with the corresponding propagators.

\section{Conclusions}
We have developed a gauge-theoretic framework for analyzing the separation of quantum subspace evolution into geometric and dynamical contributions. By constructing product representations of the Schr\"odinger propagator, we identified history-dependent dressed generators, including the {\it quasi}-holonomy connection, whose associated transport law resembles conventional holonomy, but do not arise from a genuine connection on the underlying Grassmannian bundle. The essential obstruction is gauge-theoretic: the dressed generators transform in a manner that depends explicitly on propagators encoding the past evolution, thereby violating the local transformation law of a principal-bundle connection. This shows that geometric and dynamical effects are generally inseparable, except in special commuting cases.

\section*{Acknowledgements}
E. S. acknowledges financial support from the Swedish Research Council (VR) through 
Grant No. 2025-05249.


\begin{thebibliography}{99}

\bibitem{wilczek84} F. Wilczek and A. Zee, Appearance of gauge structure in simple dynamical systems, 
Phys. Rev. Lett. {\bf 52}, 2111 (1984).

\bibitem{anandan88} J. Anandan, 
Non-adiabatic non-Abelian geometric phase, 
Phys. Lett. A {\bf 133}, 171 (1988). 

\bibitem{fredriksson26b} A. Fredriksson and E. Sj\"oqvist, 
Separation of quantum time evolution into holonomic and dynamical parts, 
Phys. Rev. A {\bf 113}, 062202 (2026). 

\bibitem{kult06} D. Kult, J. {\AA}berg, and E. Sj\"oqvist, 
Noncyclic geometric changes of quantum states, 
Phys. Rev. A {\bf 74}, 022106 (2006). 

\bibitem{chruscinski04} D. Chru\'{s}ci\'{n}ski and A. Jamio\l{}kowski, 
{\it Geometric Phases in Classical and Quantum Mechanics} 
(Birkh\"auser, Basel, 2004).

\bibitem{leone19} R. Leone, 
On the parallel transport in quantum mechanics
with an application to three-state systems, 
arXiv:1903.04928. 

\bibitem{yu23} X.-D. Yu and D. M. Tong, 
Evolution operator can always be separated into the product of holonomy and dynamic operators, 
Phys. Rev. Lett. {\bf 131}, 200202 (2023).

\bibitem{fredriksson26a} A. Fredriksson and E. Sj\"oqvist, 
Comment on ``Evolution operator can always be separated into the product of holonomy 
and dynamic operators'',
Phys. Rev. Lett. {\bf 136}, 068901 (2026). 

\end{thebibliography}
\end{document}